\documentclass{aa}  

\usepackage{graphicx}
\usepackage{txfonts}
\usepackage{subfigure}
\usepackage{epstopdf}

\begin{document} 
\titlerunning{Extra-Tidal Stars Around M22, NGC~1851 and NGC~3201}
\authorrunning{Kunder, Bono et~al.}

   \title{Spectroscopic Signatures of Extra-Tidal Stars Around the Globular Clusters NGC~6656 (M22), NGC~3201 and NGC~1851 from RAVE}


   \author{A. Kunder\inst{1},
	G. Bono\inst{2,3},
	T. Piffl\inst{4}, 
	M. Steinmetz\inst{1}, 
	E. K. Grebel\inst{5},
	B. Anguiano\inst{6,7},
	K. Freeman\inst{8}, 
	G. Kordopatis\inst{9},
	T. Zwitter\inst{10},
	R. Scholz\inst{1}, 
	B. K. Gibson\inst{11,12},
	J. Bland-Hawthorn\inst{13}, 
	G. Seabroke\inst{14}, 
	C. Boeche\inst{5},
	A. Siebert\inst{15}, 
	R.F.G.~Wyse\inst{16},
	O. Bienaym\'{e}\inst{15},
	J. Navarro\inst{17}, 
	A. Siviero\inst{18}, 
	I. Minchev\inst{1}, 
	Q. Parker\inst{6, 7, 19},  
          W. Reid\inst{6,7},
          G. Gilmore\inst{9},
	U. Munari\inst{20}
          \and
	A. Helmi\inst{21}
          }

   \institute{Leibniz-Institut f\"{u}r Astrophysics (AIP), An der Sternwarte 16, 14482 Potsdam Germany\\
              \email{akunder@aip.de}
         \and
         {Dipartimento di Fisica, Universita di Roma Tor Vergata, Rome, Italy}
         \and
	{INAF--Osservatorio Astronomico di Roma, via Frascati 33 00040 Monte Porzio Catone, Italy}
         \and
         {Rudolf Peierls Centre for Theoretical Physics, 1 Keble Road, Oxford, OX1 3NP, UK}
         \and
	{Astronomisches Rechen-Institut, Zentrum f\"{u}r Astronomie der Universit\"{a}t Heidelberg, 
	M\"{o}nchhofstr. 12-14, D-69120 Heidelberg, Germany}
         \and
         {Department of Physics \& Astronomy, Macquarie University, Sydney, NSW 2109 Australia}
         \and
	{Research Centre for Astronomy, Astrophysics and Astrophotonics, Macquarie University,
	Sydney, NSW 2109 Australia}
         \and
         {Research School of Astronomy \& Astrophysics, The Australian National University, 
         Canberra, Australia}
         \and
	{Institute of Astronomy, Cambridge University, Madingley Road, Cambridge CB3 0HA, UK}
         \and
         {University of Ljubljana, Faculty of Mathematics and Physics, Jadranska 19, 1000 Ljubljana, Slovenia}
         \and
	{Institute for Computational Astrophysics, Dept of Astronomy \& Physics, Saint Mary 
	University, Halifax, NS, BH3 3C3, Canada}
         \and
         {Jeremiah Horrocks Institute, University of Central Lancashire, Preston, PR1 2HE, United Kingdom}
         \and
	{Sydney Institute for Astronomy, School of Physics A28, University of Sydney, NSW 2006, Australia}
	\and
	{Mullard Space Science Laboratory, University College London, Holmbury St Mary, Dorking, RH5 6NT, UK}
	\and
	{Observatoire Astronomique de Strasbourg, Universit\'{e} de Strasbourg, CNRS, UMR 7550, 11 rue de l'Universit\'{e}, F-67000 Strasbourg, France}
	\and
	{Department of Physics and Astronomy, Johns Hopkins University, 3400 North Charles Street, Baltimore, MD 21218}
	\and
	{Senior CIfAR Fellow, Department of Physics and Astronomy, University of Victoria, Victoria, BC, Canada V8P 5C2}
	\and
	{Department of Physics and Astronomy, Padova University, Vicolo dell'Osservatorio 2, 35122 Padova, Italy}
	\and
	{Australian Astronomical Observatory, PO Box 915, North Ryde, NSW 1670, Australia}
	\and
	{INAF National Institute of Astrophysics, Astronomical Observatory of Padova, 36012 Asiago (VI), Italy}
	\and
	{Kapteyn Astronomical Institute, University of Groningen, P.O. Box 800, 9700 AV Groningen, The Netherlands}
	}

   \date{A\&A accepted, Aug 2014}

 
  \abstract
   {Stellar population studies of globular clusters have suggested that the brightest clusters in the Galaxy might actually be the remnant nuclei of dwarf spheroidal galaxies. If the present Galactic globular clusters formed within larger stellar systems, they are likely surrounded by extra-tidal halos and/or tails made up of stars that were tidally stripped from their parent systems.}
   {The stellar surroundings around globular clusters are therefore one of the best places to look for the remnants of an ancient dwarf galaxy.  Here an attempt is made to search for tidal debris around the supernovae (SNe) enriched globular clusters M22 and NGC~1851 as well as the kinematically unique cluster NGC~3201.}
   {The stellar parameters from the Radial Velocity Experiment (RAVE) are used to identify stars with RAVE metallicities, radial velocities and elemental-abundances consistent with the abundance patterns and properties of the stars in M22, NGC~1851 and NGC~3201.
 }
   {The discovery of RAVE stars that may be associated with M22 and NGC~1851 are reported, some of which are at projected distances of $\sim$10 degrees away from the core of these clusters.  Numerous RAVE stars associated with NGC~3201 suggest that either the tidal radius of this cluster is underestimated, or that there are some unbound stars extending a few arc minutes from the edge of the cluster's radius.  No further extra-tidal stars associated with NGC~3201 could be identified.  The bright magnitudes of the RAVE stars make them easy targets for high resolution follow-up observations, allowing an eventual further chemical tagging to solidify (or exclude) stars outside the tidal radius of the cluster as tidal debris.  In both our radial velocity histograms of the regions surrounding NGC~1851 and NGC~3201, a peak of stars at $\sim$230~km~s$^{-1}$ is seen, consistent with extended tidal debris from $\omega$~Centauri.  }
   {}

   \keywords{Galaxy: evolution -- Galaxy: formation -- Galaxy: globular clusters: individual: M22 -- 
   		   Galaxy: globular clusters: individual: NGC~1851-- Galaxy: stellar content -- Galaxy: structure 
               }

   \maketitle
%
\clearpage
\section{Introduction}
The classical Searle \& Zinn (1978) scenario of the formation of the Milky Way predicts
that the halo globular clusters (GCs) formed in larger dwarf galaxies.
Upon the accretion of smaller galaxies, dark matter, stars,
gas and in some cases GCs (Da Costa \& Armandroff 1995; Pe\~{n}arrubia, Walker \& Gilmore 2009) 
will remain in the halo of the more massive galaxy.  They
will then add to any pre-existing GC population (Searle \& Zinn 1978; Zinnecker et al. 1988; 
Freeman 1993; Abadi, Navarro \& Steinmetz 2006, Forbes \& Bridges 2010; Leaman, VandenBerg, \& Mendel 2013).

There is now good evidence that two of the most luminous GCs in the Milky Way are
the nuclei of dwarf galaxies (e.g.,~$\omega$~Centauri, Lee et al. 1999, and ~M54 
is associated with the Sagittarius dwarf spheroidal galaxy, Sarajedini \& Layden 1995).  Also, in the past two decades, 
observational studies have shown the presence of tidal tails around globular 
clusters \citep[e.g.,][]{grillmair95, grillmair96, holland97, lehmann97, leon99, leon00, testa00, odenkirchen01, odenkirchen03, sohn03, lee03, lee04, jordi10}.  
Some of these tidal tails show structures elongated for $\sim$45 degrees on the 
sky \citep[e.g., NGC~5466][]{belokurov06, grillmair06} and 
$\sim$22 degrees on the sky, \citep[e.g., Palomar 5,][]{grillmair06}.  Several other stellar 
debris features associated with globular clusters have also been 
identified, some with a possible origin from known GCs \citep[e.g., NGC~288,][]{grillmair13} 
and some associated with unknown 
clusters \citep{grillmair09, grillmair11, williams11, wyliedeboer12}, awaiting further 
and/or deeper imaging and spectroscopy to confirm such an association.

There is evidence that Galactic globulars can be split into two sub-samples,
the ones that formed in situ and those that have been accreted. The 
exact fraction is still under debate and depends on the criteria adopted 
to constrain the kinematics \citep[e.g.,][]{salaris02, marin09, vandenberg13, leaman13}. 
The two sub-samples cover 
a broad range in metal-abundance and in total mass. This means that the latter 
parameter does not allow us to discriminate between the two sub-groups. 
More recent findings suggest that the in situ GCs and the accreted ones 
do obey two different age-metallicity relations \citep{vandenberg13}. 
This evidence, once confirmed by independent analysis, might provide the 
opportunity to discriminate the origin of GCs. The above authors suggested 
that the in situ clusters are, at fixed age, $\sim$0.5~dex more metal-rich.

\citet{mackey04} utilized the horizontal branch (HB) morphology of a large sample of 
Milky Way GCs, together with GCs in dwarf spheroidal galaxies, to propose that accretion 
onto the early Milky Way was substantial, with 41 (27 per cent) of the GCs in the Milky Way 
being accreted.  Using blue halo stars, Unavane, Wyse \& Gilmore (1996) suggested an
accretion fraction closer to 10 per cent.  More recently, \citet{forbes10} 
suggest that 27 -- 47 GCs ($\sim$1/4 of the entire system), from six to eight dwarf galaxies, 
were accreted to build the Milky Way GC system we see today.
The age-metallicity relation (AMR) of Milky Way GCs further seems to support
that the metal-poor GCs were formed in relatively low-mass (dwarf) 
galaxies and later accreted by the MW \citep{leaman13}, although this is still not a
settled issue \citep[see e.g.,][]{marin09}.  

Whereas tidal tails originating from a cluster could indicate that the 
cluster is the remnant of an ancient galaxy progressively disrupted by the interaction with the 
Milky Way potential, internal processes including stellar evolution, gas expulsion, and 
two-body relaxation \citep[e.g.,][]{geyer01, kroupa01} can also dissolve a GC.  
For example, \citet{gnedin97} show how disk shocking affects each star in the cluster 
as it crosses the disk, leading to the result that stars close to the tidal boundary can 
be lost during disk crossing events given that this extra energy allows them to escape 
from the star cluster. Simulations further show that tidal features are expected for 
e.g., GCs on eccentric orbits near their apogalacticon \citep{montuori07}.   
\citet{gnedin97} find that more than half of the present Galactic GCs are to be destroyed 
within the next Hubble time.  One might expect some difference in the stellar populations
(e.g. age, elemental abundances) in the tails and main body if the
cluster were indeed a remnant of a larger system in which there were
gradients, although the extent may need to be considered on a 
case-by-case basis.

It is now accepted that most, if not all, Milky Way GCs have had at least two epochs of star 
formation \citep[e.g., see the reviews by][]{gratton04, gratton12a}.
A first epoch of star formation gave origin to the `normal' (first-generation) stars, with CNO and 
other abundances similar to Population II field stars of the same metallicity.  Afterwards, some 
other epoch of star formation (second-generation) occurred, including material heavily processed 
through the CNO cycle.  Hence, dissolved massive globular clusters can be traced by the chemical signature 
of their second-generation stars, which tend to be enhanced in light elements. 
Using such light element abundance anomalies in present-day halo field stars, 
\citet{martell10} and \citet{martell11} suggest that at least 17\% of the present-day 
stellar mass of the Milky Way halo originally formed in globular clusters.

GC streams can be some of the coldest stellar substructures yet discovered 
\citep[e.g.][]{combes99, odenkirchen09, willett09}, especially compared to early 
accretion/dissolution events which would result in low surface brightness,
dynamically hotter and spatially extended substructures.  Large samples of such 
streams can therefore provide the possibility to map the distribution of Galactic dark matter with much 
greater spatial resolution than what is presently possible \citep[e.g.,][]{murali99}.
Especially large field studies observed
in the framework of the e.g., Sloan Digitial Sky Survey \citep[SDSS;][]{york00} and 
Wide-Field Infrared Survey Explorer \citep[WISE;][]{wright10}
have led to an extensive mapping and detection of GC tidal tails.
The RAdial Velocity Experiment \citep[(RAVE)][]{steinmetz06} is one of the few kinematic
surveys that has also made discoveries of stellar streams 
associated with globular clusters \citep[e.g.,][]{williams11, wyliedeboer12}.
To date, RAVE has gathered a half-a-million 
medium resolution spectra (R$\sim$7500) of stars in the Southern sky, and
because RAVE's input catalogue is magnitude limited (8 $<$ $I$ $<$ 13), the 
stellar sample is essentially homogeneous and free of kinematic biases. 
The catalogued stars lie mostly within 2.5~kpc of the 
Sun (e.g., Burnett et al. 2011; Binney et al. 2014), although stars also belonging to the 
Large Magellanic Cloud at a distance of $\sim$50 kpc have been identified within RAVE \citep{munari09}.

In this paper, the first results from a search for extra-tidal stars around the halo GCs M22, 
NGC~1851 and NGC~3201 in the RAVE survey footprint are presented. These clusters were 
chosen for a number of reasons.  First, unlike a number of other clusters within the RAVE footprint, 
they were not specifically targeted by RAVE, so the relative frequency of stars identified within 
the cluster tidal radius and outside the cluster tidal radius can be directly compared.  Second, 
these clusters are already good candidates for having an extra-galactic origin.  Third, these 
clusters have radial velocities and/or $\rm [Fe/H]$ abundances that are largely offset from disk 
field stars seen in projection along that line of sight, reducing field star contamination issues. 
Lastly, a number of stars within these clusters have already been well-studied spectroscopically, 
so the light elements of potential extra-tidal candidates can provide important clues as to their 
cluster association. The basic properties and distances of these clusters are listed in Table~1.

We define an extra-tidal star to be an unbound star when it exceeds the tidal 
radius \citep[e.g.,][]{meylan97}, as N-body models show that this leads to consistent 
results (Giersz \& Heggie 1997).  RAVE red giant candidates in M22, NGC~1851 and 
NGC~3201, are presented, which lie outside the cluster tidal radius, but which were still likely 
born in these clusters as evidenced by their radial velocities, metallicities and 
elemental-abundances. 

%
\begin{table*}
\centering
\caption{Globular Cluster Sample} 
\label{ravegcs}
\begin{tabular}{lcccccrccccc}
Cluster Name & RA & Dec & $l$ & $b$ & $R_\odot$ & $R_{GC}$ & $\rm [Fe/H]$ & $V_{LOS}$ & $r_c$ & King $r_t$ & Wilson $r_t$  \\ 
 & (J2000) & (J2000) & ($^\circ$) & ($^\circ$) & (kpc) & (kpc) & (dex) & (km~s$^{-1}$) & (') & (') & (') \\ 
\hline
NGC~3201 &  		10 17 36.82 & $-$46 24 44.9 & 277.23 & 8.64 & 4.9 & 8.8 & $-$1.59 & 494.0 & 1.30 & 25.35 & 54.8 \\
NGC~1851 &  		05 14 06.76  & $-$40 02 47.6  & 244.51 & $-$35.03  & 12.1 & 16.6 & $-$1.18 & 320.5 & 0.09 & 6.52 & 44.7 \\
NGC~6656 (M22)    &  18 36 23.94  & $-$23 54 17.1  &   9.89   & $-$7.55 &  3.2  &  4.9  & $-$1.70 & $-$146.3 & 1.33 & 31.90 & 119.9 \\ 
\end{tabular}
\end{table*}

\begin{table*}
\centering
\caption{Globular Cluster Member Selection Criteria} 
\label{selection}
\begin{tabular}{ll}
Cluster Name & Selection Criteria \\
\hline
NGC~3201 &  	480~km~s$^{-1}$ < RV < 510~km~s$^{-1}$; RV error $<$ 10~km~s$^{-1}$  \\
NGC~1851 &  	300~km~s$^{-1}$ < RV < 380~km~s$^{-1}$; RV error $<$ 10~km~s$^{-1}$; $K > -7 \times$\hbox{ (J--K)\/} + 14  \\
NGC~6656 (M22)    &   $-$200~km~s$^{-1}$ < RV < $-$120~km~s$^{-1}$; $\rm [Fe/H] < -$1.25 dex; SNR $>$ 21; AlgoConv = 0 \\ 
\end{tabular}
\end{table*}

\section{Detection in RAVE}
The fourth data release (DR4) of RAVE contains stellar atmospheric parameters 
(effective temperature, surface gravity, overall metallicity), radial velocities, individual abundances 
and distances determined for 425~561 stars \citep{kordopatis13}.  This wealth of information 
and the extensive spatial distribution presents an ideal foundation for our study.  From this
sample, 187~305 stars have spectra for which the chemical abundance pipeline is able to 
determine abundances for magnesium, aluminum, silicon, titanium, iron and 
nickel \citep{boeche11}.  The uncertainties on these quantities vary as a function
of signal-to-noise (SNR), number of lines that could be measured, and also
depend on the specific element.  Typically, the uncertainty in $\rm [Fe/H]$ is $\sim$0.23~dex,
the uncertainty for Mg, Al, and Si is $\sim$0.2 dex, and the uncertainty for Ti and Ni
is $\sim$0.3 dex \citep[see discussion and Figures~13-17 in][for further details]{kordopatis13}.
Because the GCs NGC~3201 and NGC~1851 
have stars with radial velocities offset from the rest of the field population by at least 
one hundred km~s$^{-1}$, stars belonging to these clusters can be provisionally identified 
from their radial velocities, colors and magnitudes alone.  

Whenever possible, RAVE stellar 
abundances are used to further link extra-tidal stars to a cluster, although the 
number of RAVE spectra with reliable abundances is not nearly as extensive 
($\sim$1/4 the sample) as the RAVE sample of stars with well-determined radial velocities.  
However, it is not always obvious that the metallicity of an extra-tidal star
should be the same as a cluster star, e.g., if the metallicities of
extra-tidal stars should be the same between the GC and its dwarf galaxy host.  
For example, if a closed-box model of chemical evolution is
assumed, and assuming that the yield of each generation of stars is constant, then
the variation of metallicity ($Z$) with time can be expressed as
\begin{equation}
Z(t) = -\rho \ ln \frac{M_g(t)}{M_g(0)}
\end{equation}
\citep{binney98}.  Here, $[M_g(t) / M_g(0)]$ is the ratio of gas masses, and is unity
at $t$ = 0 and zero at the present time ($t$ = 14~Gyr).  The yield is given by $\rho$.
Setting $\rho$ to 0.005 so as to match the zero point of the Sagittarius dwarf spheroidal galaxy (Sgr)
age-metallicity relation, of which the globular cluster M54 is thought to be the nucleus \citep{layden00}, 
then $\rm [Fe/H]$ would become more metal-rich 
by $\sim$0.5 dex with every $\sim$4 Gyr that passed.  Hence, if we accept the possibility that a
nucleated dwarf galaxy, with a Milky Way GC as its nucleus, formed stars continuously in 
relative isolation with little or no infall or release of its interstellar gas, and that the yield is similar
to that of Sgr, then we would not expect a large difference in $\rm [Fe/H]$ between the GC and the extra-
tidal stars, unless we are probing extra-tidal stars with a multiple Gyr age difference from
the GC.

It is important to point out, that although a closed-box model of chemical evolution fits the age-metallicity relation of
Sgr galaxy as a whole well \citep{layden00}, this is not the case for the formation
and evolution of the anomalous GC $\omega$ Cen \citep[e.g.,][]{romano06}.  However, the $\rm [Fe/H]$
spread of the stars within $\omega$ Cen is still thought to be similar to those in
its extended tidal debris \citep{majewski12}.  Ultimately, it may be that the exact nature of the metallicity 
of an extra-tidal star linked to a host dwarf galaxy and the metallicity of a star in its nucleus (now observed as 
a Milky Way GC) depends on the particular cluster/system under study and must be considered on a 
case-by-case basis.

We also investigated using proper motions of the RAVE stars as an exclusion criterion 
in the selection of extra-tidal stars, but the errors of available measurements prevented 
the use of proper motions at this stage.  The properties of the GCs studied here are 
listed in Table~\ref{ravegcs}, taken from Harris (2010 edition of 1996), except for the 
Wilson~(1975) tidal radius, which is from McLaughlin~\&~van~der~Marel~(2005).  The columns contain 
(1) the cluster name, 
(2) the right ascension in hours, minutes and seconds (epoch J2000), 
(3) the declination in degrees, arcminutes and arcseconds, 
(4) the Galactic longitude in degrees,
(5) the Galactic latitude in degrees,
(6) the distance from Sun in kiloparsecs (kpc),
(7) the distance from Galactic center (kpc), assuming $R_0$=8.0~kpc \citep[e.g.,][]{groenewegen08, matsunaga09},
(8) the $\rm [Fe/H]$ metallicity,
(9) the Heliocentric radial velocity in km~s$^{-1}$
(10) the core radius in arcmin,
(11) the \citet{king66} tidal radius in arcmin,
and 
(11) the \citet{wilson75} tidal radius in arcmin from \citet{mclaughlin05}.

The specific selection criteria to search for cluster stars belonging to these GCs
are listed in Table~\ref{selection}.  Briefly, for NGC~3201, we only select candidates based on
their radial velocity and the errors in radial velocity measurements.  For the cluster NGC~1851, 
a color criterion is also used, and for M22, only stars with derived metallicities from RAVE are used. 
Although ideally we would prefer to select cluster stars based on their radial velocity alone, this
is only realistic for NGC~3201, the cluster with the most extreme velocity.
The AlgoConv criteria indicates which stars have reliable abundances, but as mentioned previously, 
greatly restricts the sample of RAVE stars.  Therefore, this criteria is only adopted for the cluster 
M22, the cluster with the least extreme radial velocity and for which large reddening effects makes 
selecting cluster stars by color difficult.

\subsection{NGC~3201}
NGC~3201 has long been known as a kinematically unique GC: It has the most extreme 
radial velocity of the Milky Way GCs, with a heliocentric radial velocity of 
494~km~s$^{-1}$ \citep[2010 edition of][]{harris96}.  It also has a large azimuthal velocity of 
250~km~s$^{-1}$ around the Galactic center \citep{gonzalez98}, but in a retrograde
sense.  These kinematic peculiarities have been taken as strong evidence of a possible
extra-Galactic formation with subsequent capture by the Milky Way \citep[e.g.,][]{rodgers84, vandenbergh93}.

High-resolution spectroscopy carried out on the individual stars in NGC~3201 find that despite 
its extreme kinematics, NGC~3201 shows no large spread in $\rm [Fe/H]$ compared
to the observational errors nor large chemical differences with respect to other Milky Way 
GCs \citep[e.g.][although see Simmerer et~al.~2013]{gonzalez98, carretta09n3201, munoz13}.  
Therefore, its origin appears to be similar to ``normal" GCs, i.e., with signs of multiple
populations, but without signs of supernova enrichment. 

\begin{figure*}
\centering
\subfigure{
\includegraphics[height=8.4cm]{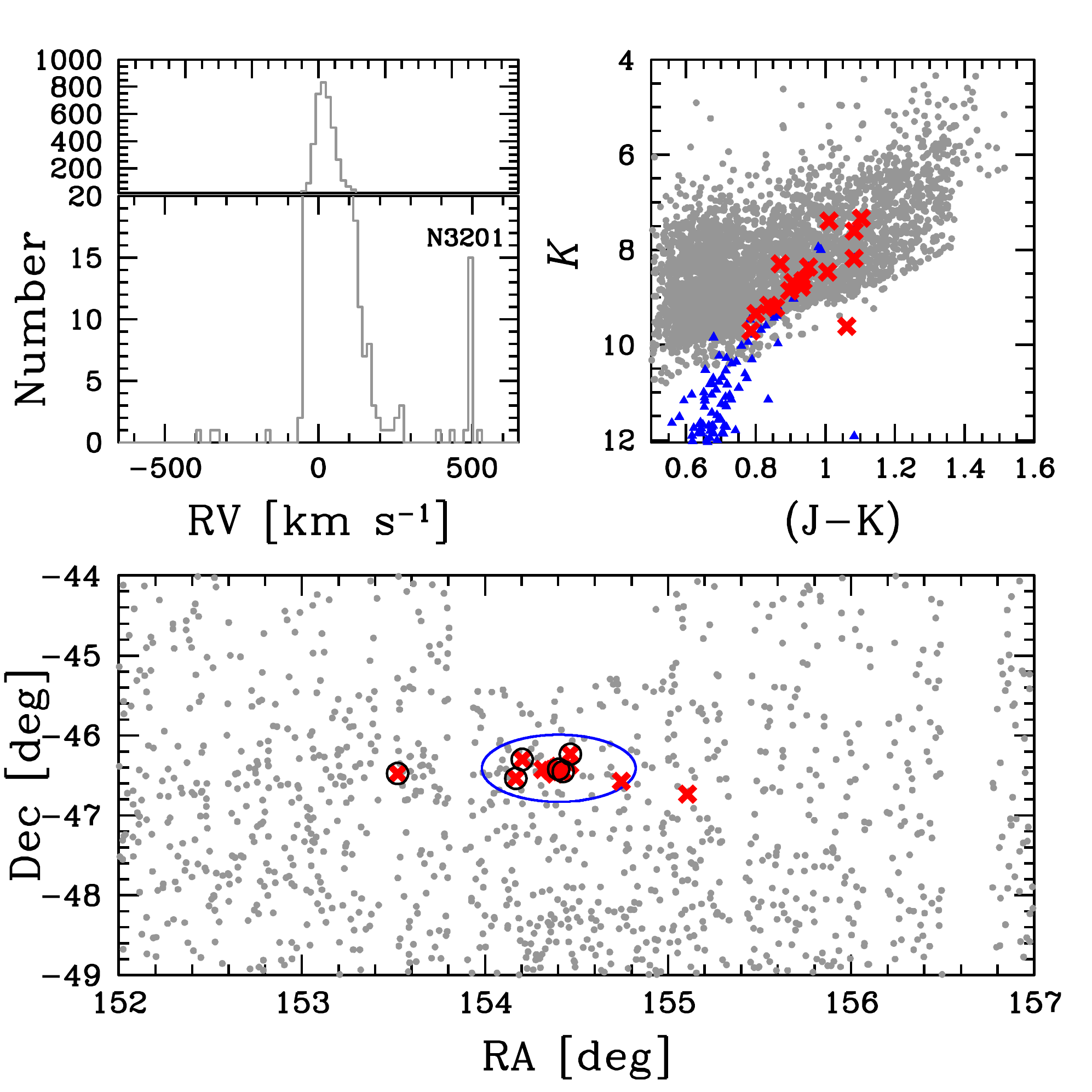}}
\quad
\subfigure{
\includegraphics[height=8.4cm]{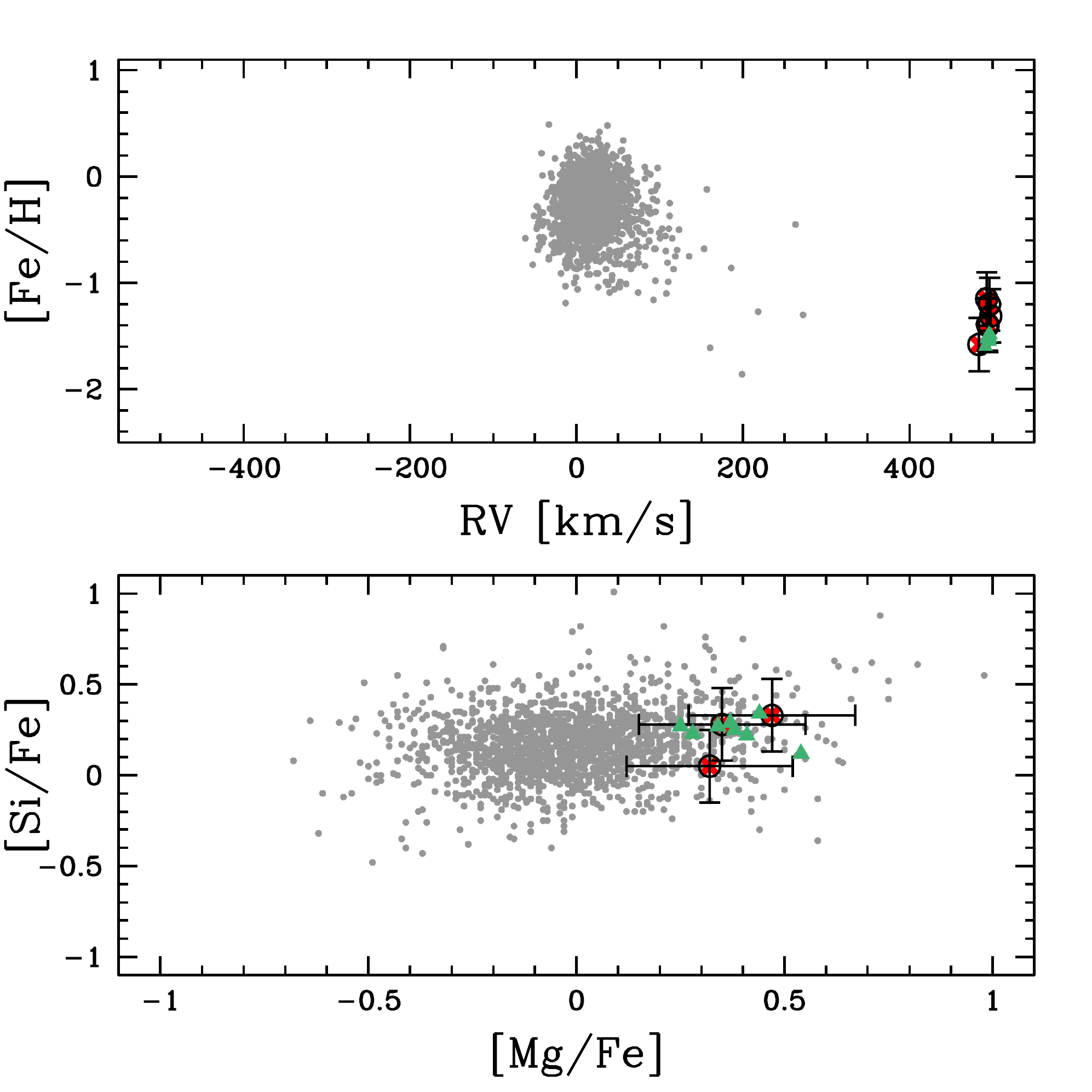}}
\caption{ {\it Left:} The CMD (top-right), and radial velocity 
distribution (top-left) of the RAVE stars in a 12 degree field centered on NGC~3201. 
The spatial distribution of the RAVE stars in a 5 degree field centered on the
cluster is shown (bottom), and the King tidal radius of the cluster is shown by the large circle.
The crosses (red in online version) show the RAVE stars whereas the triangles 
(blue in the online version) show the NGC~3201 stars studied by \citet{cote94}.
The circled RAVE stars are those for which a chemical analysis could be carried out.
{\it Right:}  The $\rm [Fe/H]$, $\rm [Si/Fe]$ and $\rm [Mg/Fe]$ ratio of a sub-sample of these same stars
for which abundances could be determined.
The triangles (green in the online version) are the parameters of the NGC~3201 red giant stars 
obtained from intermediate-resolution spectra by \citet{carretta09n3201} and high-resolution
spectra by \citet{munoz13}.
}
\label{N3201}
\end{figure*}

Figure~\ref{N3201} (left) shows the spatial location of NGC~3201 within the RAVE
footprint.  A 5 degree field of view centered on the cluster is shown. We search for 
cluster members based solely on the radial velocity
information provided by RAVE.  A radial velocity range of $\pm$15~km~s$^{-1}$ is searched, as 
this velocity range encompasses almost all stars in the extensive NGC~3201 radial 
velocity catalogue of \citet{cote94, cote95}. 
Out of the 4390 RAVE stars within a 12 degree area of the sky, 16 have a radial velocities consistent with
cluster membership (i.e., radial velocities between 480~km~s$^{-1}$ and 510~km~s$^{-1}$).
Most of these stars stars fall within the tidal radius of the cluster, but two do not (see Figure~1).   
These possible extra-tidal stars are given in Table~\ref{3201stars}, where the columns list 
(1) the RAVE ID, 
(2) the right ascension in degrees (epoch J2000), 
(3) the declination in degrees, 
(4) the line of sight radial velocity in km~s$^{-1}$,
(5) the RAVE radial velocity uncertainty in km~s$^{-1}$,
(6) the RAVE $\rm[Fe/H]$, 
(7) the RAVE $\rm[Si/Fe]$, 
(8) the RAVE $\rm[Mg/Fe]$, 
(9) the 2MASS $J$-magnitude, 
(10) the 2MASS $K$-magnitude, and 
(11) the distance from the cluster center in degrees. 
Increasing the area of the sky to cover 20 degrees of the RAVE footprint increases the number of 
RAVE stars sampled to 8495,
but no additional stars with velocities of $\sim$494~km~s$^{-1}$ are detected.  
In general, the stars with velocities of $\sim$494 km~s$^{-1}$ also have colors and magnitudes 
that place them on the giant branch of NGC~3201, further indicating their association with the cluster.
The RAVE stellar parameters are consistent with these stars being giant stars, as they all fall within
the temperature and gravity range of 4000~K $<$ $T_{eff}$ $<$ 7000~K and log~$g$ $<$ 3.5 (cgs), 
clearly separating them from e.g., nearby dwarf stars.

Six of the NGC~3201 stars have metallicities derived from the RAVE chemical abundance pipeline
and are shown in Figure~\ref{N3201} (right).  Given the external errors 
($\sim$0.23~dex) in the RAVE $\rm [Fe/H]$ metallicities as discussed previously, 
we do not wish to place too much emphasis on the metallicites to characterize the
potential extra-tidal stars.  However, it is worth stressing that the low metallicities 
of these stars set them apart from the majority
of the RAVE stars and confirm their halo membership.  Moreover, three of these stars
have spectra for which elemental abundances can be determined, all of which show enhanced $\rm [Si/Fe]$ and 
$\rm [Mg/Fe]$ values, in good agreement
with high-resolution spectroscopy of NGC~3201 from \citet{carretta09n3201} and \citet{munoz13}.


\subsection{NGC~1851}
It is traditionally considered that ``normal" GCs are those that exhibit
the following properties:
(i) their stars appear to have the same $\rm [Fe/H]$;
(ii) their stars are chemically homogeneous in the heavy elements; and
(iii) their stars are chemically inhomogeneous in the light element 
abundances, e.g., variations in the CH, CN, NH bands,
and in the O, Na, Al, and Mg abundances. Nearly all GCs
have these characteristics.  NGC~1851 is one of few GCs that
show a spread in the heavier elements and $\rm [Fe/H]$ 
\citep[rms scatter $\sim$0.05~dex in {$\rm [Fe/H]$}, which is larger than the observational errors;][]{carretta11}.
This is a characteristic that is 
seen only in a small number of other systems\footnote{The other
GCs distinguishable from normal GCs as evidenced by their spread in $\rm [Fe/H]$ are
$\omega$~Centauri \citep{lee99, bedin04}, M~54 \citep{layden97, siegel07}, Terzan~5 \citep{ferraro09},
M~22 \citep{norris83, dacosta09, marino09, marino11} and NGC~2419 \citep{cohen10, dicriscienzo11}.}.  
The presence of heavy element 
abundance ranges in these systems necessarily means their nucleosynthetic history 
must be more complicated than for ``normal" GCs, although the exact cause
remains unclear.  However, the chemical analogies with the cluster systems
$\omega$~Centauri and M~54, usually believed to be associated with the remnants of a dwarf 
galaxy cannibalized by the Milky Way, suggest the intriguing idea that all of the
clusters with $\rm [Fe/H]$ metallicity spreads could be surviving nuclei of more massive systems.
NGC~1851 is therefore a strong candidate to search for extra-tidal stars.

NGC~1851 further has a split sub-giant branch 
\citep{milone08}, a split RGB when particular filters are used \citep{han09} and spectroscopic observations 
have revealed a bimodality in $s$-process abundances \citep{yong08, villanova10, gratton12b}.  

\begin{figure*}
\centering
\subfigure{
\includegraphics[height=7.4cm]{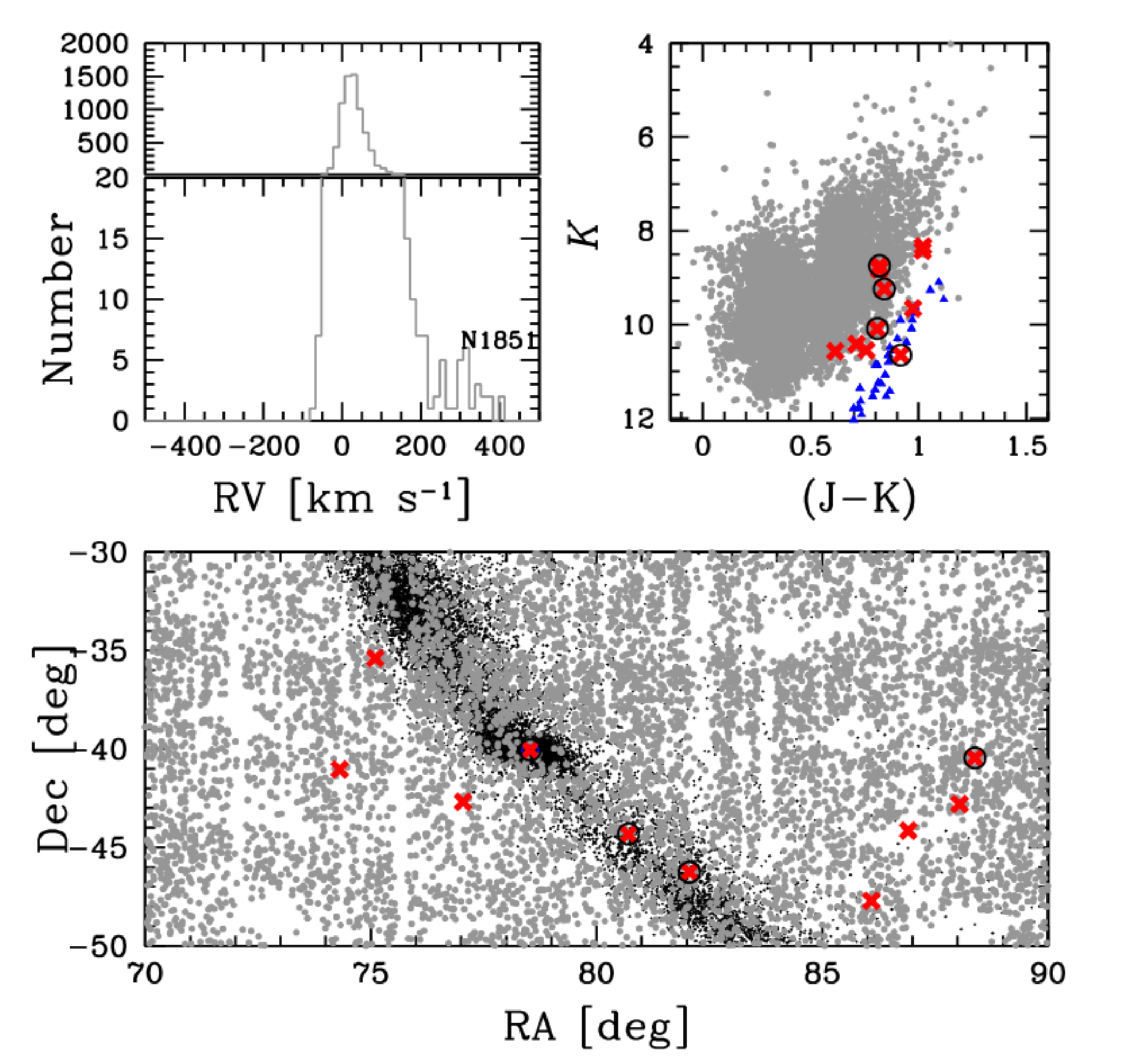}
}
\quad
\subfigure{
\includegraphics[height=7.4cm]{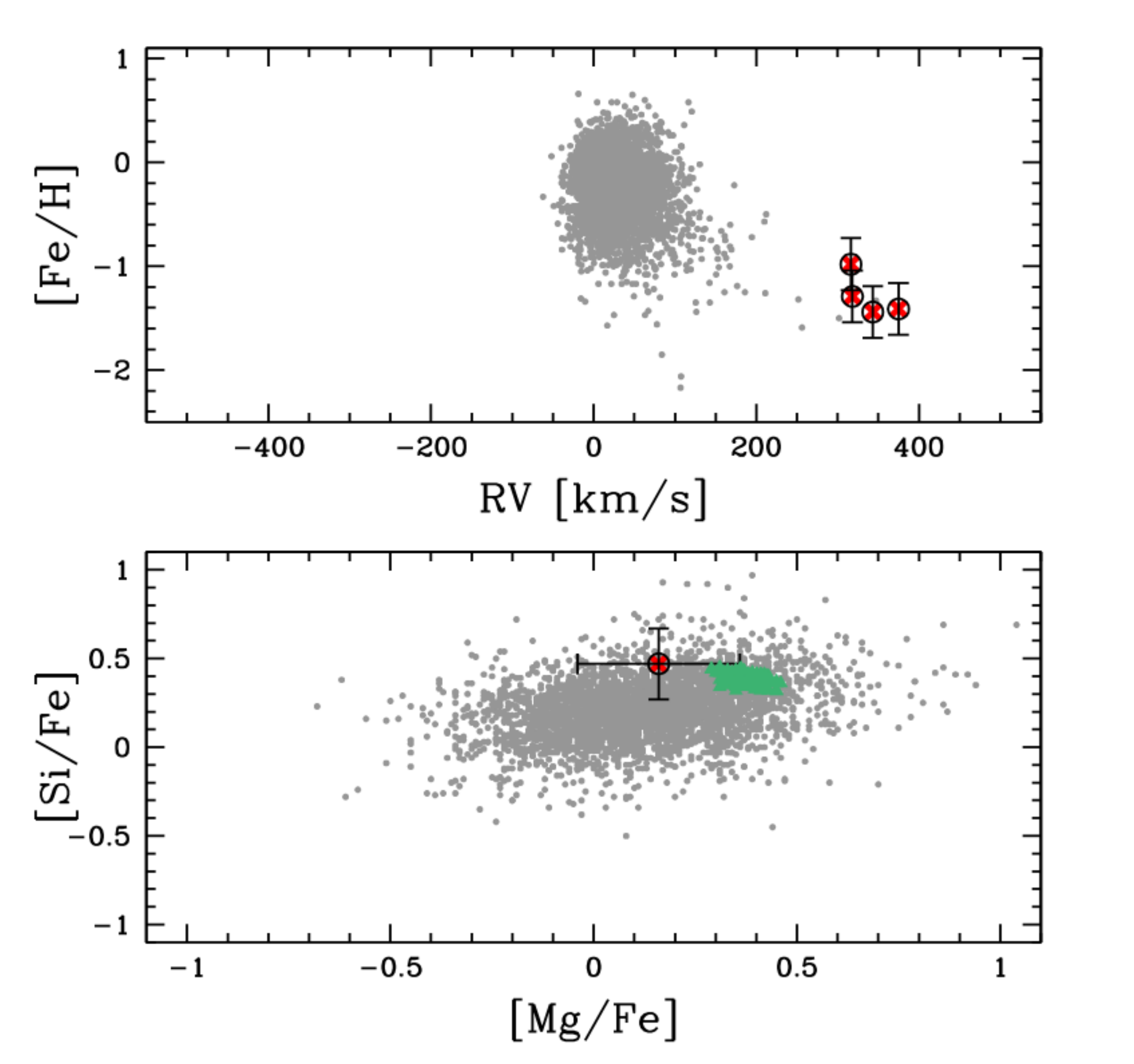}
}
\caption{ {\it Left:}  The spatial distribution (bottom), CMD (top-right), and radial velocity 
distribution of the RAVE stars in a 20 degree field centered on NGC~1851. 
The crosses (red in the online version) show the RAVE stars whereas the triangles 
(blue in the online version) show the NGC~1851 stars from \citet{milone09} and
\citet{carretta11}.
The circled RAVE stars are those for which a chemical analysis could be carried out.
In the bottom panel we also show the expected angular distribution of tidal debris stars 
(black dots) as predicted by an $N$-body simulation of the cluster orbiting the MW for one Gyr on 
the most likely orbit (see \S3.2 for details).
 {\it Right:}  The $\rm [Fe/H]$, $\rm [Si/Fe]$ and $\rm [Mg/Fe]$ ratio of a sub-sample of these same stars
for which abundances could be determined.
The triangles (green in the online version) are the parameters of the NGC~1851 red giant stars 
obtained from intermediate-resolution spectra by \citet{carretta11} and \citet{carretta12}.
}
\label{N1851}
\end{figure*}

Figure~\ref{N1851} shows the spatial location of NGC~1851 within the RAVE survey.
The central velocity dispersion is twice as large for this cluster as compared to NGC~3201 \citep{harris96}, 
so a wider range in radial velocity is searched.  We further only consider stars
as potential cluster members if their magnitudes and colors are within that of the cluster (see Table~\ref{selection}).
Here we allow for a $\sim$2~kpc distance spread in the CMD, which encompasses $\sim$95\% of the stars 
in our N-body simulation of the cluster orbiting the MW for one Gyr on the most likely orbit (see \S3.2 for details).
This also encompasses the uncertainties in the 2MASS magnitudes, which can reach up to $\sim$0.17 mag in
$K$ for the stars in the cluster core (see Table~\ref{1851stars}).

Out of the 8,948 RAVE stars surveyed in a 20 degree field around NGC~1851, eleven have
radial velocities, colors and magnitudes consistent with that of the cluster, and these stars are listed in Table~\ref{1851stars}.  
Most of these stars lie outside the tidal radius of the cluster.  The two RAVE stars
located within the tidal radius of the cluster almost certainly are cluster
members, and the nine RAVE stars which also align along the AGB and RGB of NGC~1851
are plausible cluster members, as they have properties consistent with not only the cluster's 
radial velocity, but also its distance.  Further, for the RAVE stars with a SNR 
greater than 40, the RAVE temperatures and gravities are in good agreement with the stars being giant 
branch stars of NGC~1851.

Figure~\ref{N1851} (right) shows the derived $\rm [Fe/H]$ abundances for the four RAVE stars with
spectra suitable for abundance determination.  Although the rms spread in $\rm [Fe/H]$ as determined for this cluster
from high-resolution spectra is $\sim$0.05 dex \citep{carretta11}, the individual errors in the RAVE metallicity
values ($\sim$0.23 dex) are considerably larger than this, and so it is not surprising that the $\rm [Fe/H]$ values
show a broader spread.  Within the errors, all the four stars with radial velocities consistent with that of NGC~1851 also have $\rm [Fe/H]$
metallicities in agreement with the $\rm [Fe/H]$ of the cluster and are significantly more 
metal-poor then the majority of the field population seen in projection along that line of sight in this area of the sky. 
One of these stars also has an estimate of $\rm [Si/H]$, 
and $\rm [Mg/H]$, which is consistent within its uncertainty with those of the cluster.  


\begin{table*}
\begin{scriptsize}
\centering
\caption{RAVE NGC~3201 Extra-Tidal Stars} 
\label{3201stars}
\begin{tabular}{lcccccrccccc}
RAVE ID & RA & Dec & $V_{LOS}$ & $\sigma_{V}$ & $\rm [Fe/H]$ & $\rm [Si/Fe]$ & $\rm [Mg/Fe]$ & $J$ & $K$ & r \\ 
 & (J2000) & (J2000) & (km~s$^{-1}$) & (km~s$^{-1}$) & (dex) & (dex) & (dex) & (mag) & (mag) & (degrees) \\ 
\hline
20110520\_1023m47\_056  &  153.5235  & $-$46.4779  & 495 & 1 &  $-$1.41 & 0.05 & 0.32 & 10.01$\pm$0.02 & 9.17$\pm$0.02 & 0.88 \\
20090404\_1023m47\_058  &  155.1078  & $-$46.7349  & 497 & 1 & $-$1.24 & -- & -- & 8.41$\pm$0.03 & 7.40$\pm$0.02 & 0.78 \\
\end{tabular}
\end{scriptsize}
\end{table*}

\begin{table*}
\begin{scriptsize}
\centering
\caption{RAVE NGC~1851 Stars} 
\label{1851stars}
\begin{tabular}{lcccccrccccc}
RAVE ID & RA & Dec & $V_{LOS}$ & $\sigma_{V}$ & $\rm [Fe/H]$ & $\rm [Si/Fe]$ & $\rm [Mg/Fe]$ & $J$ & $K$ & r \\ 
 & (J2000) & (J2000) & (km~s$^{-1}$) & (km~s$^{-1}$) & (dex) & (dex) & (dex) & (mag) & (mag) & (degrees) \\ 
\hline
20051110\_0449m41\_116$^b$ & 74.3136 & $-$41.0417 & 308 & 2  & -- & -- & -- & 10.63$\pm$0.02 & 9.65$\pm$0.02 & 4.33 \\
20081017\_0449m41\_116$^b$ & 74.3136 & $-$41.0417 & 305 & 1  & -- & -- & -- & 10.63$\pm$0.02 & 9.65$\pm$0.02 & 4.33 \\
20100124\_0501m36\_070 & 75.1000 & $-$35.3876 & 354 & 1  & -- & -- & -- & 9.44$\pm$0.03 & 8.42$\pm$0.02 & 5.78 \\
20111020\_0505m41\_122 & 77.0456 & $-$42.6795 & 359 & 1  & -- & -- & -- & 9.36$\pm$0.02  & 8.34$\pm$0.02 & 3.02 \\
20111020\_0505m41\_095$^a$ & 78.5297 & $-$40.0491 & 318 & 1 & $-$1.29 & -- & -- & 11.57$\pm$0.09  & 10.65$\pm$0.11  & 0.0030 \\
20061120\_0505m41\_094$^a$ & 78.5285 & $-$40.0457 & 323 & 1 & -- & -- & --  &  9.64$\pm$0.38  & 8.82$\pm$0.17  & 0.0009 \\
20120107\_0532m42\_060 & 80.7067 & $-$44.3188 & 375 & 2 &  $-$1.41 & -- & -- & 9.56$\pm$0.02  & 8.74$\pm$0.02 & 4.80 \\
20070206\_0536m47\_056$^b$ & 82.0560 & $-$46.2541 & 343 & 1  & $-$1.44 & -- & -- & 10.08$\pm$0.02 & 9.24$\pm$0.03 & 7.14 \\
20041101\_0523m48\_083$^b$ & 82.0560 & $-$46.2541 & 341 & 2  & -- & -- & -- & 10.08$\pm$0.02 & 9.24$\pm$0.03 & 7.14 \\
20050331\_0555m46\_021 & 86.0893 & $-$47.7149 & 321 & 7  & -- & -- & -- & 11.13$\pm$0.02  & 10.42$\pm$0.02 & 10.77 \\
20040209\_0555m46\_063 & 86.9080 & $-$44.1528 & 316 & 4  & -- & -- & -- & 11.18$\pm$0.02  & 10.57$\pm$0.02 & 9.33 \\
20050221\_0549m40\_146 & 88.0446 & $-$42.8051 & 311 & 1  & -- & -- & -- & 11.30$\pm$0.02  & 10.54$\pm$0.02 & 9.91 \\
20120122\_0558m41\_132 & 88.3831 & $-$40.4547 & 316 & 1 &  $-$0.98 & 0.47 & 0.16 & 10.89$\pm$0.02  & 10.09$\pm$0.02 & 9.86 \\
\hline
\hline
$^a$ within tidal radius \\
$^b$ star observed twice by RAVE \\
\end{tabular}
\end{scriptsize}
\end{table*}

\subsection{NGC~6656 -- M22}
M~22, like NGC~1851, is another one of the six GCs
known to host groups of stars with different iron $\rm [Fe/H]$
and heavy element contents.  The stellar rms scatter is $\rm[Fe/H]$$\sim$0.15 dex and 
variations in its elements associated with $s$-process elements are also 
present \citep{norris83, dacosta09, marino09, marino11}.  This direct evidence for 
extended star formation suggests this cluster may have an extra-galactic origin and hence
a tidal tail.  The CMD of M~22 shows a double sub-giant branch
\citep{marino12} as well as two discrete distributions of RGBs,
each having a different Ca abundance as observed 
from the $hk$ index of the $Ca$-$by$ photometry \citep{lee09}.

\begin{figure*}
\centering
\subfigure{
\includegraphics[height=8.4cm]{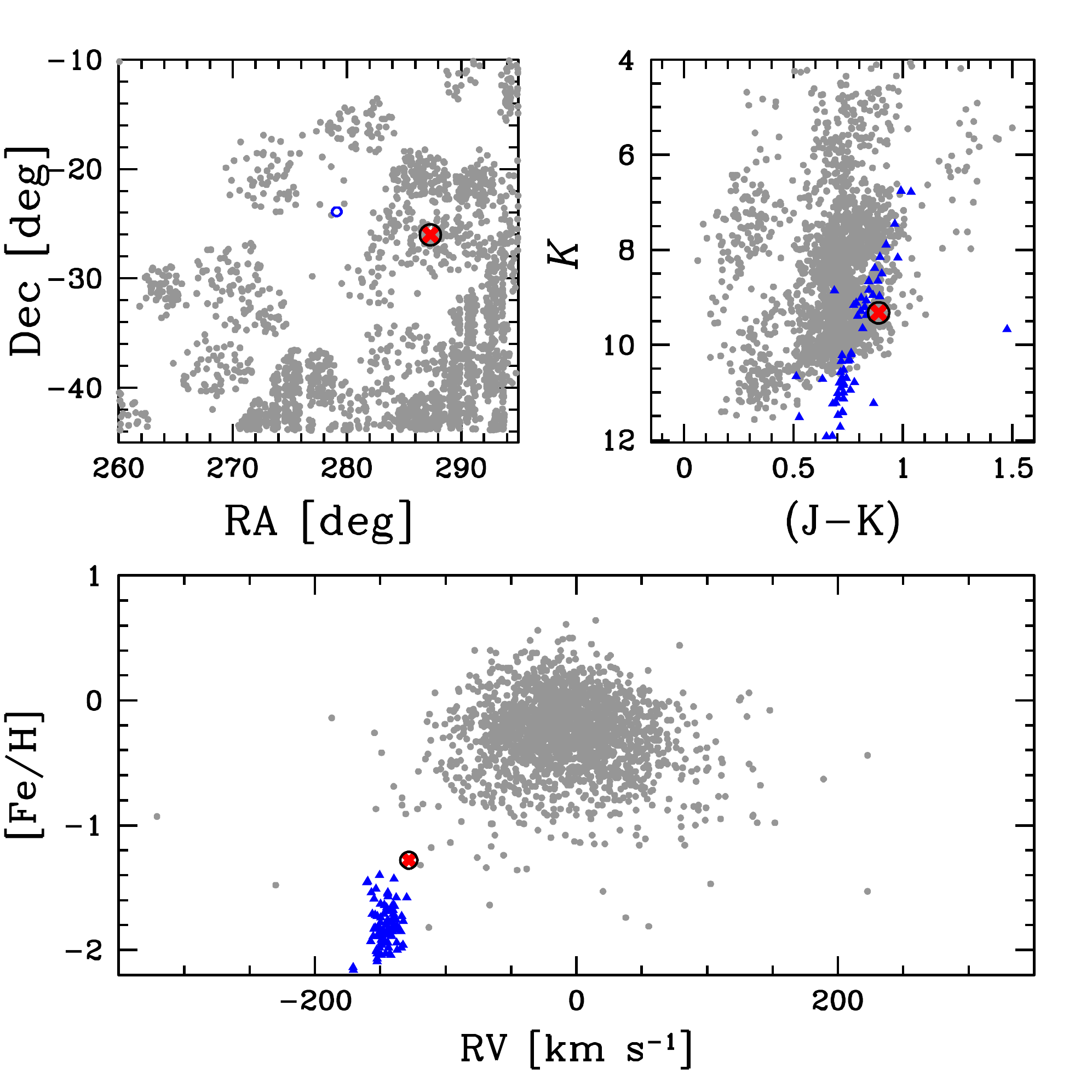}
}
\quad
\subfigure{
\includegraphics[height=8.4cm]{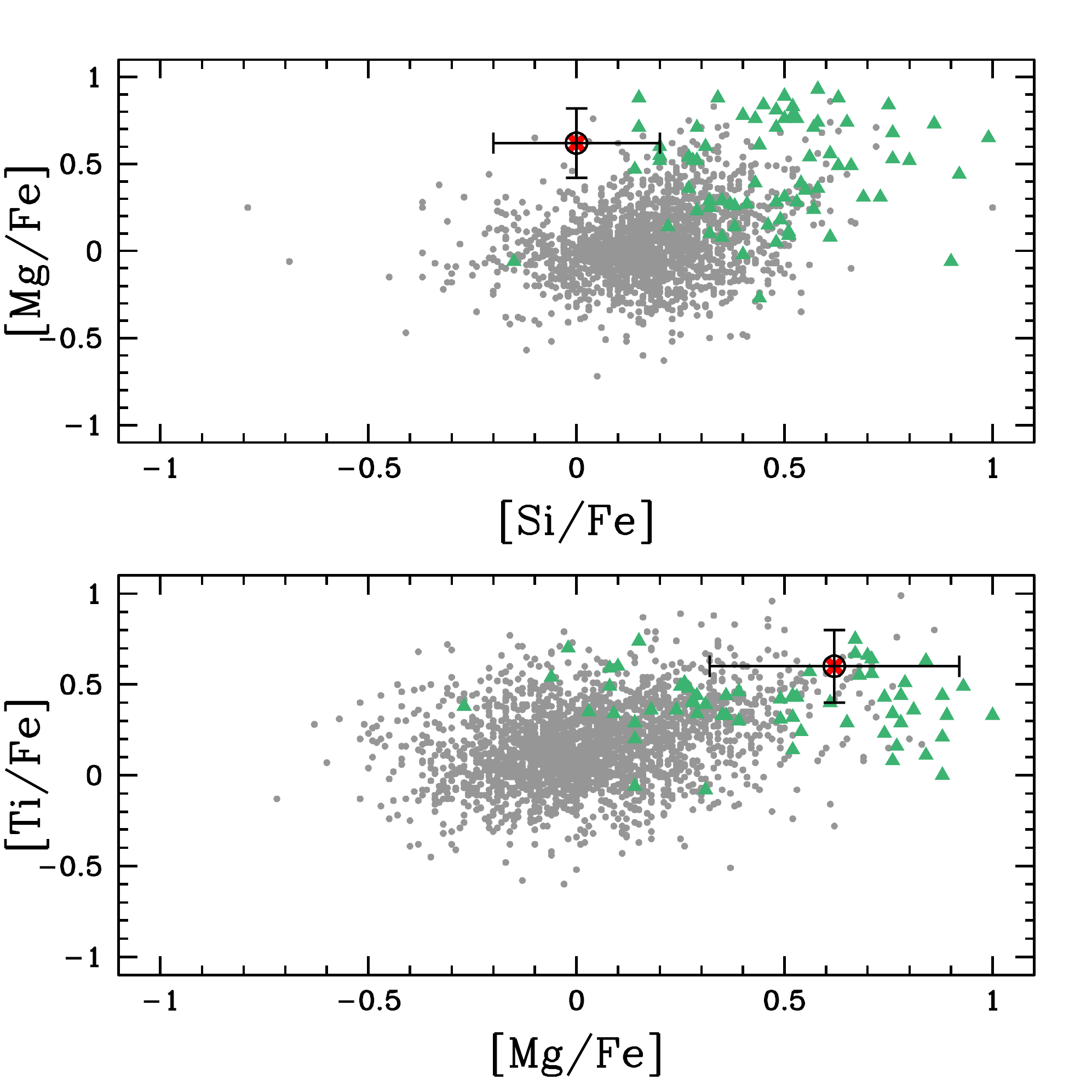}
}
\caption{ {\it Left:} The spatial distribution (top-left), CMD (top-right), and radial velocity vs $\rm [Fe/H]$
distribution of the RAVE stars in a 35 degree field centered on NGC~1851.
The cluster center of M22 is indicated by an open (blue) circle. 
The $\rm [Fe/H]$, radial velocity, colors and magnitudes of the red giant branch stars of
M22 shown as triangles (blue in the online version) are taken from \citet{marino09},
\citet{dacosta09}, \citet{alvesbrito12} and \citet{gratton14}.
 {\it Right:}  The $\rm [Si/Fe]$, $\rm [Mg/Fe]$ and $\rm [Ti/Fe]$ ratio of the same stars.
The triangles (green in the online version) are elemental abundances of M22 horizontal branch stars 
obtained from intermediate-resolution spectra by \citet{gratton14}.
}
\label{M22}
\end{figure*}

RAVE avoided probing close to the Galactic plane, so the number of RAVE stars is not as dense 
in this region of the sky as e.g., NGC~1851.  
Figure~\ref{M22} shows the spatial location of the globular cluster M22, with the 2162 RAVE stars in 
a 35 degree field of view.  Because this cluster does not have as extreme a velocity as
for NGC~1851 and NGC~3201, the RAVE stars are restricted to only those which also
have reliable abundance determinations, as indicated by sufficient SNR and
an AlgoConv value that indicates that the DR4 pipeline converged \citep[see][for further details]{kordopatis13}.
Because the stars in M22 do not have a homogenous distribution of $\rm [Fe/H]$, but have been shown to range from
$-$1.25 to $-$2.4 dex \citep[e.g.][]{gratton14}, our $\rm [Fe/H]$ search criteria cover all RAVE stars more metal poor
than $-$1.25 dex.  Figure~\ref{M22} (bottom, left panel) shows the range of $\rm [Fe/H]$ and radial velocity observed
for stars in M22, which guided our search criteria for potential extra-tidal stars belonging to this cluster.  The
exact criteria are explicitly listed in Table~\ref{selection}.

One of these stars has a radial velocity and metallicity consistent with
that of M22, and is located at a projected distance $\sim$10 degrees from the cluster center.  It also has a color and magnitude that 
places it on the giant branch of M22, as well
as $\rm [Ti/Fe]$, $\rm [Mg/Fe]$ and $\rm [Si/Fe]$ values consistent with those of the cluster.


\begin{table*}
\begin{scriptsize}
\centering
\caption{RAVE NGC~6656 Extra-Tidal Stars} 
\label{6656stars}
\begin{tabular}{lcccccrccccc}
RAVE ID & RA & Dec & $V_{LOS}$ & $\sigma_{V}$ & $\rm [Fe/H]$ & $\rm [Si/Fe]$ & $\rm [Ti/Fe]$ & $\rm [Al/Fe]$  & $\rm [Mg/Fe]$ & $J$ & $K$ \\ 
 & (J2000) & (J2000) & (km~s$^{-1}$) & (km~s$^{-1}$) & (dex) & (dex) & (dex) & (dex) & (dex) & (mag) & (mag) \\ 

\hline
20110824\_1915m27\_053 & 287.3046 & $-$26.0189 & $-$128 & 1 & $-$1.28 & 0.00 & 0.60 & 0.26 & 0.62 & 10.21$\pm$0.02 & 9.32$\pm$0.03 \\
\end{tabular}
\end{scriptsize}
\end{table*}

\section{Discussion}
A cluster's tidal radius is normally estimated by fitting \citet{king66} models to cluster density profiles.  
However, the assumption of the King edge radius as a real physical limit of the cluster is not
set in stone.  For example, \citet{mclaughlin05} investigate the stellar distributions of 
GCs in not only the Galaxy, but also the Magellanic Clouds and the
Fornax dwarf spheroidal galaxy, and find that \citet{wilson75} models, originally developed for 
application to elliptical galaxies, better describe the outer structures
than do King models.  More recent investigations capitalizing on wide-field imagers
confirms this result \citep[e.g.,][]{dicecco13}.  

Also, \citet{zocchi12} show with their sample of 13 GCs, that 
kinematic fits are crucial to assess if a model is actually suited to describe a given globular cluster. 
Unfortunately, there is a general lack of good kinematic data, generally consisting of a small number 
of data-points and not well distributed in radius.  The case for NGC~3201 is encouraging in that
\citet{cote95} obtained 857 radial velocities with median precision $\sim$1~km~s$^{-1}$ for 399 
member giants to trace the velocity dispersion profile of this cluster out to 32'.1.  
This has allowed nonparametric modelling of this cluster, and it was found that
NGC~3201 has significantly more stars which are tightly bound (low energy) than King models
predict \citep{gebhardt95}.

\subsection{NGC~3201}
For NGC~3201, the tidal radius from the Wilson model is 
about twice as large as from the King model (see Table~1).
The two RAVE extra-tidal stars are located 0.78$^\circ$ and 0.88$^\circ$
away in projection from the cluster center, well beyond the 0.47$^\circ$ King tidal 
radius of the cluster \citep[][2010 edition]{harris96}.  
Hence, either our results extend the velocity dispersion profile of this cluster out to
$\sim$0.9$^\circ$, inline with the \citet{gebhardt95} findings and a Wilson tidal radius, or 
these RAVE stars are unbound and are therefore not part of the cluster's tidal radius.  
It is worth noting that NGC~3201 is thought to have experienced a recent impact with the Galactic 
disk \citep{vandeputte09}, and this crossing could show stellar debris structure observed as 
clumpy structures \citep[e.g.,][]{chen10}.  The extra-tidal RAVE stars may also be associated with
just recently experienced disk-crossing shocks.

Curiously, despite the relatively rich number of RAVE cluster members
found within the cluster tidal radius of NGC~3201, further than 0.9$^\circ$, not one RAVE star
with a radial velocity within 490~km~s$^{-1}$ and 500~km~s$^{-1}$ is found when searching 64,206 
RAVE stars within a 30$^\circ$ radius of the cluster center (between the right ascension of 
124.5$^\circ$ and 184.5$^\circ$ and the declination  of $-$76.5$^\circ$ and $-$16.5$^\circ$).
This suggests that if NGC~3201 has an extra-galactic origin, no prominent, large-scale stellar debris 
consistient with radial velocities between 490~km~s$^{-1}$ -- 500~km~s$^{-1}$ exists, such as 
would be perhaps expected if it were a satellite galaxy merging with the Milky Way. 
However, \citet{cote95} found a difference in the radial velocity of 1.22$\pm$0.25~km~s$^{-1}$ for stars 
on either side of the cluster, and although this could be interpreted as the cluster having internal
rotation, it could also be that the radial velocity of the stellar population associated with NGC~3201 
has a radial velocity gradient.  There is one RAVE star outside the tidal radius of NGC~3201
with a radial velocity between 470~km~s$^{-1}$ -- 490~km~s$^{-1}$ and five RAVE stars
with radial velocities between 500~km~s$^{-1}$ -- 520~km~s$^{-1}$, all located more than 20$^\circ$
away from the cluster center.  Therefore, the RAVE radial velocity data suggest that any potential tidal
stream surrounding NGC~3201 is a weak feature.   If this GC was accreted by the Milky Way much earlier, 
any debris could be quite diffuse and 
dynamically hot \citep[e.g.,][]{bullock05, font08}.  Detailed dynamical modelling of this cluster 
could address this further.

\subsection{NGC~1851}
There is increasing speculation that NGC~1851 is the result of a merger of two GCs that
were both formed in a dwarf spheroidal galaxy, dragged to the center of the dSph by dynamical friction,
and that now this dwarf spheroidal is being destroyed by
Galactic tidal forces \citep[e.g.,][]{vandenbergh96, carretta10}.  
NGC~1851 has an unusually high central concentration of light, which can be
explained if it was formed by the merging of two initially spherical star clusters \citep[e.g.,][]{white78}.
Each cluster might have formed with a slightly different metallicity and with a 
different level of $\alpha$\-elements, which would explain that the stars in NGC~1851
can be divided into two groups according to their metallicity, where {\sl both} components show a 
moderate Na-O anticorrelation \citep{carretta10}.
Since the Na-O (and the C-N) anticorrelations alone can be considered 
as the signature of multiple stellar populations, the two stellar groups in NGC~1851 appear to 
both exhibit the product of multiple stellar formation episodes. 

That NGC~1851 was once part of a larger system, such as a  dwarf spheroidal galaxy, 
is supported by its unexpected dispersion in {\sl heavy} element abundances, which
suggests enrichment by supernovae \citep{lee09, han09}.  It is hard to understand how a 
GC with little or no dark matter and low binding energy 
could retain energetic SNe ejecta, unless the GC we observe today were a remnant of an 
initially much more massive stellar system. 

Further, a diffuse stellar halo with a size of more than 500 pc and a mass of $\sim$ 0.1 per cent of the dynamical
mass of NGC~1851 has been observed by \citet{olszewski09}, and this peculiar feature can be explained by
its formation in the central region of a defunct host dwarf galaxy \citep{bekki12}.  
The different density distributions of metal-poor and metal-rich stars suggests that if NGC~1851
was formed from the merging of two clusters, this can not have occurred too long in the past (i.e, 
less than a few Gyr ago).  This merging should have happened before the putative dwarf spheroidal 
dissolved, so it should still be possible to identify the remains of this ancestral, parent galaxy.

Given the state of published models, it is not possible to easily determine if the extra-tidal stars
detected in RAVE are following the cluster's orbit, as is the case with Pal~5 and NGC~5466.  
The RAVE observations of candidate extra-tidal stars are also limited spatially, and we therefore can not trace out discrete
structures in common with photographic counts in e.g., studies that find extensions of the
cluster \citep{leon00} or studies that show a smooth extended halo structure \citep{olszewski09}.
It is, however, noteworthy that the extra-tidal candidate stars have a preferential direction 
towards a S-E extension, similar to the tails observed around NGC~1851 from \citet{leon00},
see their Figure~10.  

To obtain some intuition about where we should expect to find stars that were stripped off the cluster, 
we performed a simple (collisionless) $N$-body experiment\footnote{The $N$-body integration was 
done using the code gyrfalcON \citep{dehnen02} which is publicly available in the $N$-body code frame 
work NEMO by \citet{teuben95}.}.  We set-up a King model with the structural properties of NGC~1851 as 
reported by Harris (1996) on an orbit around the Milky Way that would lead the cluster after $\sim$1 Gyr to 
its present position and kinematics (taken from Harris 1996 and Dinescu et~al.\ 1997). For the Galactic gravitational 
potential we use the mass model proposed by \citet{mcmillan11}. The resulting final angular positions of 
star particles are also shown in Figure~\ref{N1851}.  As we start with the current structure of the cluster, 
which is then tidally stripped along its orbit, the final cluster in our model has slightly different properties than 
NGC~1851, but the simulation is still effective to identify the rough location where the tidal debris would be observed.

The correlation between our NGC~1851 orbit and the
candidate NGC~1851 extra-tidal RAVE stars is encouraging, and indicates a low probability that the 
majority of these stars are due to random fluctuations in the field.  With suitable follow-up observations,
they could be used as probes of a debris stream associated with NGC~1851. 

Recently, a spectroscopic survey performed in the outskirts of NGC~1851 by \citet{sollima12}
resulted in tentative evidence of a cold peak in the distribution of radial velocities at $\sim$180~km~s$^{-1}$.
These stars had a location in the color-magnitude diagram compatible with a stream at a similar distance to this cluster, and 
if confirmed, would constitute a strong indication of the presence of a stream in the direction of NGC~1851.
Therefore, we searched for an excess of RAVE stars with velocities of $\sim$180~km~s$^{-1}$ within 5 degrees of the cluster. 
Only one star with a radial velocity between 170~km~s$^{-1}$ and 190~km~s$^{-1}$ was identified with a similar
location in the color-magnitude diagram as NGC~1851, located $\sim$4.5$^\circ$ from the cluster center.
Hence, the RAVE data are not able to confirm a 180~km~s$^{-1}$ stream in the direction of NGC~1851, although given that
the \citet{sollima12} feature was constituted mainly by main-sequence stars and the RAVE stars would
only be able to probe the RGB and AGB stars, this is perhaps not surprising.

\subsection{NGC~6656 -- M22}
As discussed previously, the abundance spread in heavy elements in the stars of M22 have caused
much speculation that this cluster was previously a stellar nucleus
formed in situ or through merging of two GCs with different chemical abundances.  Despite
being located in relatively close proximity to the Sun, $\rm R_{\sun}$ = 3.2 kpc, studies
of this cluster are hampered by its location toward
the crowded and heavily extincted Galactic bulge, making it difficult to disentangle cluster 
parameters, due to e.g., contamination from field stars and differential reddening \citep{monaco04, kunder13}. 

The RAVE stars are limited in the direction of M22, yet we find a candidate extra-tidal
star associated with M22 based on $\rm [Fe/H]$ metallicity, elemental-abundances and radial velocity,
as well as its position on the RGB of M22.
Similarly, within the ARGOS spectroscopic survey of the red clump stars of the Galactic bulge, 
\citet{ness13} find 11 stars with $\rm [Fe/H]$ 
abundances and radial velocities consistent with the GC M22.
They find a mean distance spread of 1.8~kpc and speculate that this distance spread is due in part
to extended tidal streams associated with M22.  Unfortunately due to the unavailability of the ARGOS data and
extremely limited data tables, we are unable to assess where these stars are spatially distributed and
hence how far they are from the cluster center.  The relative success of RAVE
and ARGOS to kinematically select M22 stars beyond its tidal radius suggests that a systematic 
spectroscopic search
for M22 extra-tidal stars would likely yield interesting results and perhaps a large enough sample to investigate
the geometry of potential tidal tails associated with this cluster.

\subsection{Comments Regarding Omega Cen}
The globular cluster $\omega$ Centauri (NGC~5129) is the largest globular cluster known in the Milky Way galaxy
with a radial velocity of $\sim$231~km~s$^{-1}$ and a large central velocity dispersion of 
$\sim$17~km~s$^{-1}$ \citep{harris96}.  The stars within this cluster show an unusually large star-to-star 
spread in iron that spans more than an order of magnitude, from $-$2.2 < {\rm [Fe/H]} < $-$0.7
\citep[e.g.,][]{johnson10}.  As already mentioned in the introduction, there has been evidence 
presented that this GC is the former nucleus of a dwarf galaxy \citep[e.g.,][]{lee99, bekki03, majewski12}.

It is worth noting that there is a clear group of 14 stars with radial velocities $\sim$220-280~km~s$^{-1}$ as seen in 
the top, left panel of Figure~\ref{N3201}, and a group of 11 stars with radial velocities $\sim$220-280~km~s$^{-1}$ in 
the top, left panel of Figure~\ref{N1851}.  The majority of these stars for which abundance determinations
could be determined are also relatively metal-poor, as shown in the top, right panel of the same figures.  These features
are consistent with stars from the massive globular cluster $\omega$~Cen, which has been shown to have
a significant kinematically coherent ``tidal debris" signature spanning $>$60$^\circ$ of Galactic longitude in the
Grid Giant Star Survey (GGSS) radial velocity survey of giant stars located within $\sim$5 kpc of the 
Sun \citep{majewski12}.  This $\omega$~Cen
signature is especially prominent at $l$$\sim$285$^\circ$ \citep{majewski12}, and coincidentally, 
NGC~3201 is located at $l$$\sim$277$^\circ$ and NGC~1851 is located at $l$$\sim$244$^\circ$.
Our radial velocity histograms covering the large areas surrounding NGC~1851 and NGC~3201 therefore 
agree with the result from \citet{majewski12}, that extended parts of the $\omega$~Cen tidal stream 
are contributing to giant stars in the inner Galaxy.  It is also in agreement with the study of
\citet{wyliedeboer10}, who find that a number of members of the Kapteyn group of nearby halo stars
are probably remnants of tidal debris stripped from the parent galaxy of $\omega$~Cen, or from the cluster itself, 
during its merger with the Galaxy.

\section{Conclusions}
Stars selected from the extensive RAVE survey that can be associated with the GCs M22, NGC~1851
and NGC~3201 are presented, some reaching outside the \citet{king66} tidal radius of these clusters.
For the kinematically peculiar cluster NGC~3201, we find cluster member stars extending few arc minutes 
from the edge of the cluster's radius and no further extra-tidal stars associated with NGC~3201.
Given the relatively rich number of RAVE cluster members
found within the cluster tidal radius of NGC~3201, the RAVE radial velocity data suggests that any potential tidal
stream surrounding NGC~3201 would be a weak feature.

For M22 and NGC~1851, two GCs with groups of stars with different $s$-element content, each group 
exhibiting their own Na-O, C-N anticorrelations, we find RAVE stars with radial velocities,
$\rm [Fe/H]$ abundances, elemental-abundances and positions on the CMD consistent with 
that of the cluster, but located at projected distances of $\sim$10 degrees from the cluster center.  Although the stellar proper motion 
errors of these stars are too large to use as an inclusion criteria, the UCAC4 proper motions are consistent 
with the small mean absolute proper motions of the corresponding clusters taking into account the errors. 
We conclude that these
stars are promising extra-tidal candidates, suggesting an extra-galactic origin for these clusters.

Since the tidal tails of globular clusters are preferentially formed by the lowest-mass stars 
\citep[e.g.][]{combes99, koch04}, 
it is challenging to study tidal tails of globular clusters using only bright stars.  
Therefore, it is difficult to use the RAVE results presented here to comment on specifics of
accreted GCs in the Milky Way, although if these extra-tidal stars are shown to be
extended structures originating from the GC, then this provides spectroscopic evidence 
that accretion onto the early MW was significant, as predicted in the seminal work of \citet{searle78}.
Following up with deep CCD studies in the location of the sky of these extra-tidal stars to search
for the cluster's main-sequence, as in e.g., \citet{carballobello10}, 
would be a promising next step to
confirm the association of these RAVE stars with the remnants of an accreted dwarf galaxy
in the disk of the Milky Way.

\section{Acknowledgments}
The authors thank the reviewer for her/his useful comments that helped improve
and clarify this manuscript. 
This work was partially supported by PRIN--INAF 2011 ``Tracing the
formation and evolution of the Galactic halo with VST" (P.I.: M. Marconi)
and by PRIN--MIUR (2010LY5N2T) ``Chemical and dynamical evolution of
the Milky Way and Local Group galaxies" (P.I.: F. Matteucci).
Funding for RAVE has been provided by: the Australian Astronomical Observatory; 
the Leibniz-Institut fuer Astrophysik Potsdam (AIP); the Australian National University; 
the Australian Research Council; the French National Research Agency; the German 
Research Foundation (SPP 1177 and SFB 881); the European Research Council 
(ERC-StG 240271 Galactica); the Istituto Nazionale di Astrofisica at Padova; The 
Johns Hopkins University; the National Science Foundation of the USA (AST-0908326); 
the W. M. Keck foundation; the Macquarie University; the Netherlands Research School 
for Astronomy; the Natural Sciences and Engineering Research Council of Canada; the 
Slovenian Research Agency; the Swiss National Science Foundation; the 
Science \& Technology Facilities Council of the UK; Opticon; Strasbourg Observatory; 
and the Universities of Groningen, Heidelberg and Sydney. The RAVE web site is at http://www.rave-survey.org 


\end{document}